\documentclass[12pt,preprint]{aastex}
\usepackage{natbib}
\received{}
\accepted{}
\author{
David E. Johnston,\altaffilmark{1}
Erin S. Sheldon,\altaffilmark{2,3}
Argyro Tasitsiomi,\altaffilmark{2,3}
Joshua A. Frieman,\altaffilmark{2,3,4}
Risa H. Wechsler,\altaffilmark{2,3,5}
Timothy A. McKay\altaffilmark{6}
}
\altaffiltext{1}{Princeton University Observatory, Peyton Hall, Princeton, NJ 08544. 
email davej@astro.princeton.edu}
\altaffiltext{2}{Kavli Institute for Cosmological Physics, The University of 
  Chicago, 5640 South Ellis Avenue, Chicago, IL 60637.}
\altaffiltext{3}{Department of Astronomy and Astrophysics, The University of
  Chicago, 5640 South Ellis Avenue, Chicago, IL 60637.}
\altaffiltext{4}{Fermi National Accelerator Laboratory, P.O. Box 500, Batavia,
  IL 60510.}
\altaffiltext{5}{Hubble Fellow, Enrico Fermi Fellow}
\altaffiltext{6}{Physics Department, University of Michigan, AnnArbor, MI 48109.}
\begin{document}

\title{Cross-Correlation Lensing: Determining Galaxy and Cluster  
Mass Profiles from Statistical Weak Lensing Measurements}

\begin{abstract}
We present a new non-parametric method for determining mean 3D density and
mass profiles from weak lensing measurements around stacked samples of
galaxies or clusters, that is, from measurement of the galaxy-shear or 
cluster-shear correlation functions. Since the correlation function is 
statistically isotropic, this method evades problems, such as projection of 
large-scale structure along the line of sight or halo asphericity, 
that complicate attempts to infer masses from weak 
lensing measurements of individual objects. We demonstrate the
utility of this method in measuring halo profiles, galaxy-mass and
cluster-mass cross-correlation functions, and cluster virial masses. We
test this method on an N-body simulation and show that it correctly and 
accurately recovers the 3D density and mass profiles of halos. We find 
no evidence of problems due to a mass sheet degeneracy in the simulation. 
Cross-correlation lensing provides a powerful method for calibrating 
the mass-observable relation for use in measurement of the
cluster mass function in large surveys. 
It can also be used on large scales to measure
and remove the halo bias and thereby provide a direct measurement of
$\Omega_m\sigma_8$.  
\end{abstract}


\section{Introduction}
Gravitational lensing has become a powerful tool for studying
cosmology and especially for studying the unseen matter which
dominates the Universe. Weak gravitational lensing 
produces an image of the Universe which is
slightly distorted --- these distortions reveal the quantity and
clustering properties of the elusive dark matter. One major
application of weak lensing has been measurement of the mass distribution
around rich galaxy clusters
\citep{fahlman:cluster-lensing,tyson-fischer:1689,luppino-kaiser:cluster,
clowe-luppino:cluster,joffre:lensing-3667,dahle:weak-lensing-clusters,cypriano:24-xray}
This has allowed the reconstruction of 2D projected cluster mass density maps 
which, combined with optical and X-ray measurements, has enabled
measurement of mass-to-light ratios and determination of the cluster 
baryon fraction. 

Another important use of weak lensing is the
measurement of the power spectrum of mass fluctuations through cosmic
shear, which has now been detected by several groups
\citep{van-waerbeke:cosmic-shear,wittman:cosmic-shear,bacon:cosmic-shear,maoli:cosmic-shear,
rhodes:cosmic-shear-groth,hoekstra:cosmic-shear,jarvis:cosmic-shear,bacon-massey:cosmic-shear,
hamana:cosmic-shear}.
Cosmic shear, or shear-shear correlations, probes cosmology through its dependence on both the
growth of structure and the expansion history of the Universe
\citep{hu:cs-params,abazajian:cs-params,takada-jain:cs-params,song-knox:cs-params,refregier:wl-space-params}
since it is sensitive only 
to the matter distribution, cosmic shear is theoretically clean. 

While cluster lensing allows the study of the mass distribution of
individual objects, cosmic shear is a purely statistical quantity
analogous to the angular power spectrum of the cosmic microwave background. 
There is another method, called galaxy-galaxy lensing or galaxy-shear 
correlations, which
is a hybrid of the two. Galaxy-galaxy lensing measures the mean weak
lensing shear produced by a sample of lens galaxies by averaging the
shear signal over large numbers of them. It is thus a statistical
measure which can be applied to different samples of galaxies to
measure the average mass distribution around different galaxy types.
After early null results
\citep{kristian:gal-gal,tyson:gal-gal}, the first detections using
this technique were made by
\cite{brainerd:gal-gal,dellantonio:gal-gal,griffiths:gal-gal}; these
measurements have quickly improved
\citep{hudson:gal-gal,fischer:gal-gal,wilson:gal-gal,mckay:gal-gal,smith:gal-gal,hoekstra:gal-gal-cnoc}.
The Sloan Digital Sky Survey (SDSS) \citep{york:sdss} has provided an
excellent sample for measuring galaxy-galaxy lensing:
\cite{sheldon:gmcf} measured the weak lensing signal around the
SDSS main spectroscopic sample to 
high accuracy (see also \citep{seljak:bias}). The galaxy-galaxy
lensing signal is best interpreted, in the language of large-scale
structure, as an estimate of 
the galaxy-mass correlation function. Combined with the
galaxy auto-correlation function, it gives us information about galaxy
biasing; on large scales, it provides a direct measurement of
$\sigma_8 \Omega_m$ that assumes only the physics of general
relativity.

Another arena in which such statistical 
weak lensing methods can be applied is the study 
of galaxy clusters. The number counts of galaxy clusters as a function
of mass and redshift is a sensitive probe of cosmological parameters. The
cluster mass function has been measured by several groups using various
proxies for cluster mass
\citep{bahcall-cen:clusters,white-ef-frenk:clusters,viana-liddle:clusters,
reiprich:clusters,bahcall:sdss-clusters}. The
measurement of cluster number counts at low redshift is usually 
used to constrain
the combination $\sigma_8 \Omega_m^{\beta}$ with $\beta \sim 0.5$, although the exact value of
$\beta$ really depends on which range of mass scales is being probed \citep{rozo}.
Measurement of the evolution of the mass function with redshift allows one to
break the degeneracy between  these two parameters 
\citep{bahcall-bode:ev-mass-func}. Galaxy clusters
are strongly biased compared to the dark matter, and this bias as a function
of halo mass is a prediction of theoretical models
\citep{kaiser:cluster-bias, mo-white:cluster-bias, sheth-tormen:cluster-bias,
seljak-warren:halo-bias}.  A measurement of cluster bias as a function of
mass via cluster correlations thus provides an independent constraint on 
$\sigma_8$ \citep{rozo}. In the future, planned large-area surveys 
will aim to determine the properties of the dark energy (its equation of 
state and energy density) by counting clusters and measuring their 
clustering to redshifts $z \sim 1-2$ \citep{haimanholdermohr}. 

One of the main challenges to using galaxy clusters to constrain
cosmology is uncertainty in the determination of clusters masses.
Typically one selects clusters and/or infers their masses based on 
some observable proxy for mass, such as X-ray temperature or luminosity 
\citep{borgani-guzzo:xray-clusters,del-popolo:xray-clusters}, 
Sunyaev-Zel'dovich effect (SZE) flux decrement \citep{grego:sz-clusters}, 
optical richness, or galaxy  
velocity dispersion \citep{vaderMarel:cnoc-clusters,mckay:dynamical-conf,lokas-mamon:coma-dynamics}. 
In all these cases, 
the mass-observable relation, as well as its scatter and possible 
evolution with redshift, must be calibrated  
\citep{pierpaoli:mass-richness} either by simulations, by combining number 
counts with spatial clustering information \citep{majmohr,limahu1,limahu2}, 
or by 
`direct' mass measurements of a subsample of the clusters. X-ray and SZE 
measurements are sensitive to varying degrees to the physics of 
the intra-cluster medium, such as radiative
transfer, cooling, accretion, and feedback mechanisms associated with
star and galaxy formation, possibly leading to scatter in the mass-observable 
relation.  Dynamical mass measurements through the galaxy velocity dispersion 
suffer from uncertainties in the anisotropy of the velocity distribution 
function, possible bias of the galaxy velocities relative to the dark 
matter, and possible violation of the required assumption of dynamical
equilibrium (e.g., due to recent mergers).

Weak lensing provides a method for measuring
cluster masses directly and thereby calibrating mass-observable relations. 
The advantage of lensing is that it is
sensitive to the total mass regardless of its makeup or dynamical
state; the only physics involved is the deflection law given by
general relativity. However, lensing measurements have
limitations of their own.  Weak lensing mass reconstruction of individual 
clusters can only obtain a
two-dimensional surface mass density; information about the structure along
the line of sight is irretrievably lost.  Typically these measurements
come from pointed observations of known massive clusters, selected 
by X-ray luminosity or other property. Because the
observations need to go deep enough to obtain the necessary density of
background source galaxies to make spatially resolved mass maps, 
they are time-consuming and therefore tend to be 
limited in area, usually not extending much beyond 
the virial radius.  Because of this limited range of physical
scales covered, there are uncertainties arising from the mass sheet
degeneracy \citep{bradac:mass-sheet}. In addition, large-scale structure along
the line of sight and chance projections with clusters at different redshifts 
can influence individual cluster mass determinations, causing $\sim 30$\% 
scatter around the true mass and possibly a bias in the mass determination 
\citep{cen:projection-effects,metzler:cluster-lss-contam,hoekstra:lss-contam1,metzler:lss-contam,
white:clusters-completeness,hoekstra:lss-contam,clowe:lensing-proj,dodelson:lss-contam}.

In this paper, we demonstrate a new way to measure the average density
and mass profile of clusters using stacked samples that, in principle, can be 
selected by any observable proxy.  We introduce a
completely non-parametric method to deproject the average shear profile and
obtain the mean 3D density and mass profiles for a sample.  This method
only assumes the statistical 
isotropy of the cluster-mass correlation function. We
perform tests on an N-body simulation to show that the method
correctly recovers the average 3D density and mass profiles. From the
derived mass profile of the clusters, one can measure a virial radius
and virial mass without assuming a model for the profile. Unlike
individual cluster lensing reconstructions, these mass estimates are
not affected by large-scale structure or chance projections 
along the line of sight. We show
that these methods will allow one to constrain the density profiles of
galaxy clusters, calibrate mass-observable relations, and thereby 
measure the cluster mass function and bias.

In \S 2, we derive the relations between tangential shear measurements 
and the 2D surface mass density. In \S 3, we invert the relation to 
obtain the 3D density profile. In \S 4, we present methods of deriving 
3D mass profiles, and in \S 5 we describe some of the practical steps 
(interpolation, extrapolation, and error propagation) needed to implement 
the method. In \S 6, we apply the method to a large N-body simulation, 
showing that it accurately recovers the correct 3D density and mass 
profiles statistically, and in \S 7 we discuss why the method is 
insensitive to asphericity of halos and projected large-scale structure 
along the line of sight. \S 8 presents our conclusions and outlook for 
the future. In future papers, we will apply this method to measurement 
of the cluster-mass correlation function for optically selected clusters 
in the SDSS and in N-body simulations, derive scaling relations 
between optical richness and virial mass, and constrain the bias 
and the mass power spectrum by comparison with the cluster-cluster 
correlation function.

\section{2D Density Inversion}

The primary measurement in cross-correlation lensing is of
the average tangential distortion of background source galaxy images by
foreground lens objects, typically galaxies or clusters. 
This distortion is measured in several
radial annuli around the center of the lens and repeated for all lenses in 
the sample. For the purpose of this paper, we assume
that we have redshift measurements for the lenses, 
as is the case with the SDSS and
some other current and planned surveys. We will also assume that we have an 
unbiased estimate of the lensing strength, $<\Sigma_{crit}^{-1}>$, which is to say we have a good
understanding of the redshift distribution of the source galaxies, possibly from
using photometric redshifts.

For any set of lenses, we can define the lens-mass correlation
function, $\xi_{lm}$, and write the average 3D mass density of the lens population as
\begin{equation}
\rho(r) = \bar{\rho}~[1+\xi_{lm}(r)] = \bar{\rho}+\Delta\rho(r), 
\label{eq:rhoofr}
\end{equation}
where the cosmic mass density 
$\bar{\rho} = \Omega_m \rho_{crit}$ and $\Delta\rho(r) \equiv
\bar{\rho}~\xi_{lm}(r)$.  The lensing deflection and the shear are
determined by the excess mass density projected along the line of 
sight, $\Sigma(R)=\int dz
\Delta\rho(r)$, where we have dropped the constant term for reasons
described below. Here and below, $R$ denotes the 2D 
separation from the lens center in the lens plane, 
$r$ denotes the 3D radial separation from the lens,
and $z$ is the distance from the lens along the line of sight, so that
$r^2=R^2+z^2$.  The average tangential shear, $\gamma_T$, in a thin annulus of 
radius $R$ is given by
\begin{equation}
\Sigma_{crit}~\gamma_T(R)=\bar{\Sigma}(<R) - \Sigma(R) \equiv \Delta\Sigma(R),
\label{eq:delta-sigma}
\end{equation}
where $\bar{\Sigma}(<R)$ is the mean surface density interior to radius $R$, 
the critical surface density is 
\begin{equation} 
\Sigma_{crit} = \frac{c^2}{4 \pi G} \frac{D_s}{D_l D_{ls}},
\end{equation}
and $D_s$, $D_{l}$, and $D_{ls}$ are the angular
diameter distances to the source, to the lens, and from lens to source. 
Defining the convergence 
$\kappa(R) \equiv \Sigma(R)/\Sigma_{crit}$, the expression for the shear 
can be put in the dimensionless form
\begin{equation}
\gamma_T(R)=\bar{\kappa}(<R) - \kappa(R). 
\label{eq:delta-kappa}
\end{equation}
Equation \ref{eq:delta-sigma} defines $\Delta\Sigma(R)$,
the difference between the mean surface density inside a disk of radius $R$
and $\Sigma(R)$ (azimuthally averaged), which is the primary
observable in cross-correlation lensing.  From this definition, 
one can see that a uniform-density 
mass sheet of transverse size $>R$ 
produces no shear; this is referred to as the
mass sheet degeneracy \citep{bradac:mass-sheet} and is the reason 
why we dropped
the constant term in the definition of $\Sigma(R)$ above.

We have shown how to predict the lensing observable $\Delta\Sigma$
from the theoretical quantity of interest, the 
correlation function $\Delta\rho=\Omega_m\xi_{lm}$. We would like
to invert this relation, to infer 
$\Omega_m\xi_{lm}$ from measurements of $\Delta\Sigma$.  One way to
proceed would be to use a parametric model for $\xi_{lm}$, such as a
power-law or NFW profile \citep{nfw:profile}.  If a model profile is
assumed, one can project the model to 2D, compute $\Delta\Sigma(R)$, 
fit the shear data, and determine the best-fit model 
parameters. This method has the advantages of 
simplicity and of small formal error bars on the derived model parameters.
Of course, the problem with any model-dependent method is that an
incorrect model will result in incorrect interpretation of the
best-fit parameters.  In addition, parametric methods force smoothness on
the data and thus do not allow one to see any features in $\xi_{lm}$ 
or detect fine-scale deviations from the model. 
Since we would like to constrain the full density profile and
not just determine its best-fit parameters for a particular 
model, in this paper
we develop a non-parametric approach which inverts the process above and
allows a direct estimate of the 3D density profile $\Delta\rho(R)$.
We start by deriving a 2D reconstruction for the mean convergence 
profile, $\kappa(R)$.

Taking the derivative of Equation \ref{eq:delta-kappa} with respect to
$R$ one can show that
\begin{equation}
-\kappa^{\prime}(R)=\gamma^{\prime}_T(R) + 2 \gamma_T(R)/R,
\label{eq:kappa-prime}
\end{equation}
where ${}^{\prime}$ denotes $d/dR$. 
This is just the azimuthally symmetric version of Kaiser's equation \citep{kaiser:nonlinear}
\begin{equation} 
\nabla \kappa = \left (  \begin{array}{c}
        \gamma_{1,1} + \gamma_{2,1} \\ \gamma_{2,1} - \gamma_{1,2} \label{eq:kaiser-del}
        \end{array} \right ),
\label{eq:delkappa}
\end{equation}
where $\gamma_i$ denotes the two-component shear field, and $\gamma_{i,j} = 
\partial \gamma_i/\partial \theta_j$ denotes the derivative in the lens plane.
Equation \ref{eq:delkappa} 
is the starting point for most finite-field non-linear inversion
methods for weak lensing
\citep{kaiser:nonlinear,lombardi-bertin:inversion,seitz-schneider:inversion}.
Note that these equations relate the shear to the density in a
\emph{local} way, in contrast to Equation \ref{eq:delta-kappa} for the
tangential shear, which is non-local.  As we will show below,
both the local and non-local aspects of these equations are useful
when inferring the mass profile.

\cite{schneider-seitz:nonlinear} pointed out that, except in the limit
of very weak fields, $\gamma, \kappa \ll 1$, 
the shear, $\gamma$, is not directly observable.
The average shape of background objects is not determined by 
$\gamma$ or $\kappa$
but rather by the degenerate combination 
$g \equiv \gamma/(1-\kappa)$, which is sometimes
referred to as the reduced shear. 
Furthermore, the average ellipticities of background
galaxies is equal to the complex distortion, $\delta
\equiv 2 g /(1+|g|^2)$. Solving this quadratic equation for $g$
results in $g=(1 \pm \sqrt{1-|\delta|^2})/\delta^{*}$. Since there are
two roots to this equation, one cannot tell from purely 
local measurements which
root to choose. This reflects an inherent local degeneracy to any weak
lensing measurements. For weak fields, 
one should take the negative root, for which $\gamma \simeq g
\simeq \delta/2$. The choice of sign is related to image parity; it 
changes when crossing critical curves in the lens plane, where the
magnification diverges and arcs are present. 
In practice, it is only in the
cores of massive clusters where one might 
make the wrong choice of parity, 
and this is unlikely to present a major problem for 
the statistical method of weak lensing we pursue here.
These rare, extreme regions are not the places 
where weak lensing will be the most useful probe of the mass distribution.

The other continuous degeneracy is not as problematic: one can show that if $g$ is
measured, one can invert for $\kappa$ up to the mass sheet
degeneracy \citep{kaiser:nonlinear}.  For our azimuthally symmetric
case, we can simply substitute $g (1-\kappa)$ for $\gamma$ in Equation
\ref{eq:delta-kappa}, resulting in
\begin{equation}
-\frac{\kappa^{\prime}}{1-\kappa} = \frac{d}{dR} \ln (1-\kappa) = 
\frac{g^{\prime} + 2 g/R}{1-g} \equiv G(R) \label{eq:non-lin-inv},
\end{equation}
so that $\ln (1-\kappa)$ is determined up to a constant. Imposing the
boundary condition $\kappa(\infty)=0$, one can integrate this ordinary
differential equation:
\begin{equation}
\kappa(R) = 1- \exp(-\int_{R}^{\infty} dp~G(p)),
\label{eq:non-lin-solution}
\end{equation}
which in the weak-field limit becomes 
\begin{equation}
\kappa(R) = \int_{R}^{\infty}dp~\left[ \gamma_T(p)^{\prime} + 2 \gamma_T(p)/p \right].
\label{eq:kappa-weak}
\end{equation}
In section \ref{section:practical}, we will discuss the truncation of
integrals of this kind to a finite region. Since Equation \ref{eq:kappa-weak} 
is linear in $\gamma_T$, one could simply multiply it by
$\Sigma_{crit}$ to obtain an expression for
$\Sigma(R)$. This is suitable for an analysis such as that in 
\cite{sheldon:gmcf}, where $\Delta\Sigma \equiv \Sigma_{crit} \gamma$
is averaged. Beyond the weak field regime, the situation 
is more complicated, because
of the $1-g$ term in the denominator of $G(R)$; in this case,
an average of the lensing distortion does not simply yield an 
average of $\Sigma(R)$. Nonetheless, as noted above, this should only be a
problem for small scales and high densities, i.e., for the cores of 
very massive clusters, 
and in that regime 
one could apply a correction or an iterative process to determine $\Sigma(R)$. 
Moreover, there is only a small 
spatial range where this correction is important and where
one can still safely assume that $\gamma = g \simeq \delta/2$ is the
correct root. The method of averaging shear profiles may simply not be
the most prudent method for probing the non-weak regime; strong lensing can
likely tell us more about the density profiles of massive clusters at
such small scales. Alternatively, other methods, such as the maximum likelihood approach 
of \cite{schneider-rix:max-like}, might be able to better handle this regime, 
though at the cost of introducing explicit model dependence.

\section{Inverting to 3D Density Profiles}

We can rewrite Equation \ref{eq:non-lin-inv} (after multiplying by $\Sigma_{crit}$) as 
\begin{equation}
-\Sigma^{\prime}(R) = \Sigma_{crit}~(1-\kappa) G(R) =
 \Sigma_{crit}~G(R)~\exp(-\int_{R}^{\infty} dp~G(p))
\end{equation}
or in terms of the shear, (multiplying Equation \ref{eq:kappa-prime} by $\Sigma_{crit}$)
\begin{equation}
-\Sigma^{\prime}(R) = \Delta\Sigma^{\prime}(R) + \frac{2}{R} \Delta\Sigma(R)
\label{eq:delta-sigma-prime}
\end{equation}

An equation for $\Sigma^{\prime}(R)$ is useful because of the
existence of the Von Zeipel (or sometimes called Abel) inversion
formula for the 3D density profile:
\begin{equation}
\Delta\rho(r) = \frac{1}{\pi} \int_r^{\infty}dR~ \frac{-\Sigma^{\prime}(R)}{\sqrt{R^2-r^2}}.
\label{eq:delta-rho}
\end{equation}
The inverse of this equation is just the usual projection
\begin{equation}
\Sigma(R) = 2 \int_R^{\infty}dr~r~\frac{\Delta\rho(r)}{\sqrt{r^2-R^2}}, 
\label{eq:sigma-proj}
\end{equation}
where, as before, $\Delta\rho(r) \equiv \rho(r)-\bar{\rho}$, because we do not 
recover the average mass density of the Universe:
lensing is not sensitive to mass sheets. That is, we have 
arbitrarily chosen the boundary condition $\rho(\infty)=\bar{\rho}$ or $\Delta\rho(\infty)=0$; for further discussion of the mass sheet degeneracy, see 
\S 5.2. 

Inversions of this type have a long history in
astronomy. \cite{vonzeipel:deproj} derived this inversion formula and
used it to determine the 3D density profile of globular clusters from
imaging data. Von Zeipel's proof of this inversion formula rests on
reducing it through substitutions to Abel's formula
\citep{abel:deproj}, which in turn, is usually proven through the use
of the Laplace transform. \cite{plummer:deproj}, also concerned with
globular clusters, derived a similar formula which uses intensities in
long parallel strips rather than circular annuli.  Recently
\cite{kaastra:deproj} has shown that Von Zeipel's formula can be
derived from Plummer's formula much more easily without using
Abel's formula.  The interested reader should consult
\cite{kaastra:deproj} as well as \cite{bremer:deproj} and references
therein. These inversions have been used for astronomical deprojections 
in many different contexts, including globular clusters, galaxy
clusters, elliptical galaxies, supernova remnants, planetary nebulae,
and galaxy correlation functions.

This inversion formula assumes spherical symmetry.  For stacked
samples of lenses, spherical symmetry follows simply from the statistical 
isotropy of the Universe. Since we are computing an average mass density of 
many lenses, it can be written as a correlation function,
$\Delta\rho(r)=\bar{\rho}~\xi_{lm}(r)$.  
Equation \ref{eq:delta-rho} combined with Equation \ref{eq:delta-sigma-prime} 
results in the Equation used in \cite{sheldon:gmcf} to measure the
galaxy-mass correlation function:
\begin{equation}
\Delta\rho(r)=\Omega_m\rho_{crit}~\xi_{lm} 
= \frac{1}{\pi}\int_r^{\infty}dR~ \frac{\Delta\Sigma^{\prime}(R) + 2 \Delta\Sigma(R)/R}{\sqrt{R^2-r^2}}.
\end{equation}
Again, one must truncate this integral since one only has measurements
out to some finite projected separation $R_{max}$. 
We discuss this truncation in Section
\ref{section:practical}.

\section{From Density to Mass Profiles}

Now that we can reconstruct mean 3D density profiles, it would seem
straightforward to calculate the 3D mass profiles, $M(r)=4\pi \int_0^r
dy~y^2 \Delta\rho(y)$. (This mass, the second moment of the
correlation function, is sometimes referred to as $J_3(r)$ in the
large-scale structure literature, though not usually with
cross-correlation functions in mind.) However, the density profile
cannot simply be integrated directly, since one only has reliable density
information from shear measurements 
down to some minimum scale, $R_{min}$. One could
extrapolate the derived density profile to smaller scales, but that 
would introduce model-dependent assumptions. As
we will show, one can do better: the mass inside $R_{min}$ 
is in fact constrained by the lensing data. This is because Equations
\ref{eq:delta-sigma} and \ref{eq:delta-kappa} for the tangential shear
are non-local. To illustrate the point, consider that a 
point mass creates a shear, $\gamma(r)\sim
r^{-2}$. However, substituting this power-law expression 
into Equation \ref{eq:kappa-prime} results in a
complete cancellation. Thus, adding a point mass does not alter the
derived 2D and 3D densities, as should be expected.  The mass inside
$R_{min}$ is in fact the piece of information that we lost
when taking the derivative of Equation \ref{eq:delta-kappa} to get
the local Equation \ref{eq:kappa-prime}.  Hence, any mass estimator should use
both the local and non-local equations.

Throughout this calculation, we will assume that the mean density of
the universe has been subtracted out, as above.  
This is just a statement of boundary
conditions, which does not affect the final 3D profiles,
and allows us to ignore the effects of the lensing
kernel $\Sigma_{crit}$.
We begin with Equation \ref{eq:delta-sigma}
\begin{eqnarray}
\Delta\Sigma(R) & =& \bar{\Sigma}(<R) - \Sigma(R) \\
 & = & \frac{m_{cyl}(R)}{\pi R^2} - \Sigma(R),
\end{eqnarray}
which defines the mass, $m_{cyl}(R)$, inside a cylinder of radius $R$ 
oriented along the line of sight:
\begin{equation}
m_{cyl}(R) = \pi R^2 [ \Delta\Sigma(R) + \Sigma(R) ]. 
\end{equation}
Using Equation \ref{eq:sigma-proj} to replace $\Sigma(R)$, we have 
\begin{equation}
m_{cyl}(R) = \pi R^2 \left[ \Delta\Sigma(R) + 2 \int_R^{\infty}dr~r~\frac{\Delta\rho(r)}{\sqrt{r^2-R^2}} \right]
\label{eq:m2d}
\end{equation}

Now we want to relate this cylindrical mass, $m_{cyl}(R)$, to the 3D spherical
mass $M(R)$.  It is a slight abuse of notation to write $M(R)$, since
$R$ is the 2D radius in the plane of the sky, but this avoids the
bulkiness of the more precise notation, $M(r=R)$. 
Again, assuming spherical symmetry,
the 3D mass can be written as the total mass inside the cylinder minus
the mass inside the cylinder but outside the sphere. Using
cylindrical-polar coordinates $x,z,\phi$ and spherical radius
$r=\sqrt{x^2+z^2}$, with the $z$-axis being the axis of projection, we have 
\begin{eqnarray}
M(R) & = & m_{cyl}(R) - 2 \pi \int_0^R dx~x~2 \int_{\sqrt{R^2-x^2}}^{\infty} dz~\Delta\rho(r) \\   
& = & m_{cyl}(R) - 4 \pi \int_R^{\infty} dr~r~ \Delta\rho(r) \left( r- \sqrt{r^2-R^2} \right).
\label{eq:MR}
\end{eqnarray}
Substituting Equation \ref{eq:m2d} for $m_{cyl}$ into \ref{eq:MR} gives an
expression for $M(R)$ in terms of $\Delta\Sigma$ and
$\Delta\rho$. Since $\Delta\rho$ is a functional of $\Delta\Sigma$, this
is a mapping: $\Delta\Sigma \longmapsto M$.  We will define this
particular form for $M(R)$ as $M_{out}(R)$, since it only uses data outside 
of $R$:
\begin{equation}
M_{out}(R) = \pi R^2 \Delta\Sigma(R) + 2 \pi \int_R^{\infty} dr~r \Delta\rho(r)
\left[ \frac{R^2}{\sqrt{r^2-R^2}} - 2 \left( r - \sqrt{r^2-R^2} \right) \right].
\label{eq:MoutR}
\end{equation}
Since we assumed above that the average
density is zero, equation \ref{eq:MoutR} actually gives only the mass
over-density, not the total mass; to properly correct for this, a
term $4/3 \pi R^3 \bar{\rho}$ should be added, but this formula is
typically used on scales where this extra contribution is negligible.

We will now derive another formula for the mass profile. As we
mentioned, the simplest idea is just to integrate the density. We will
call this particular form $M_{in}(R)$, which is given by
\begin{eqnarray}
M_{in}(R) & = & 4 \pi \int_0^R dr~r^2~\Delta\rho(r) \\
& = & M(R_{min}) + 4 \pi \int_{R_{min}}^R dr~r^2~\Delta\rho(r).
\end{eqnarray}
Replacing the first term with expression \ref{eq:MoutR} evaluated 
for $M_{out}(R_{min})$ gives
\begin{equation}
M_{in}(R) = \pi R_{min}^2 \Delta\Sigma(R_{min}) + 2 \pi \int_{R_{min}}^R dr~r~\Delta\rho(r)
\frac{2 r^2 - R_{min}^2}{\sqrt{r^2-R_{min}^2}} + E(R,R_{min}),
\end{equation}
where the last term is
\begin{equation}
E(R,R_{min}) = 2 \pi \int_R^{\infty} dr~r~\Delta\rho(r) 
\left[ \frac{R_{min}^2}{\sqrt{r^2-R_{min}^2}} - 2 \left( r - \sqrt{r^2-R_{min}^2} \right) \right].
\end{equation}
For $R \gg R_{min}$ the term $E(R,R_{min})$ is very small
compared to the rest of the terms.  Note that although we have referred
to this expression as $M_{in}$, the last term does depend on 
$\Delta\Sigma$ outside of $R$.
There are thus two seemingly different expressions for the 3D mass
profile. In fact, one could replace $R_{min}$ in the above expression
with any radius between $R_{min}$ and $R$:  all of these expressions
are equivalent, and they do not add any extra information by
combining them. They simply add up the information from different
radii in a different order, much like changing the order of
integration on a 2D integral.

For clusters, one can measure a mean virial mass from these mass profiles
quite easily. The
virial mass is defined as the mass within the virial radius,
$r_{vir}$, where the virial radius is the radius at which the
average enclosed density is $\Delta_{vir}$ times either the
average or critical density of the Universe. 
Often, $\Delta_{vir}=200$ (and critical) is used and this virial radius is often denoted $r_{200}$.  
One must then interpolate between points to find the
radius at which $M(r_{200})=4\pi/3~r_{200}^3~200~\rho_{crit}$. These
non-parametric virial masses should prove very useful in comparing
data to N-body simulations, which make frequent use of the virial mass.

As mentioned above, 
the mass profiles derived from $\Delta\Sigma$ actually contain more
information than the density profiles. The mass profile at $R_{min}$
tells us about the average density at $R < R_{min}$, so it 
can be used to probe these smaller scales. This might be useful in
answering the question of whether halos have cusps or cores at small
radii. It is equally true that $\Delta\Sigma$ contains all of the
information. However projection and non-locality make it somewhat
harder to use, except through parametric modeling. If one just wants
to fit a model to the data, one might as well use
$\Delta\Sigma$. For visualizing the actual shape of the density
profiles without a model and obtaining virial mass estimates, these
inversions are most useful.

Finally, we note that 
there has been some work on deriving aperture mass estimates in 2D
from the shear profile
\citep{schneider:aperture-masses,schneider-bartelmann:moments}. 
The aperture mass $m = \int d^2x ~\kappa(x)~
w(|x|)$, where $w$ is an azimuthally symmetric weight function.   
\cite{schneider:aperture-masses} suggests choosing a normalized 
weight function satisfying 
$\int d^2x ~ w(|x|)=0$, i.e., a compensated filter, so that
the mass estimate is not affected by the mass sheet degeneracy. Obviously this
would restrict the choice of weight functions so that, for example, 
the top-hat weight function cannot be
used. We have implicitly 
restricted ourselves above to top-hat apertures and derived a 3D
mass profile from the lensing measurements. We discuss the mass
sheet degeneracy by showing (in Section \ref{section:endpoint}) that,
for stacked samples, this unknown constant can be predicted fairly
well and only creates real uncertainty at the very largest scales.

\section{Practical Application}
\label{section:practical}

With expressions in hand for inverting 3D density and mass profiles
from the lensing data, $\Delta\Sigma$, we turn to discuss the
application of these formulae to real data. The two main issues
to consider are interpolation to evaluate the various integrals and
handling the parts of the integrals that extend beyond the scales for
which we have measurements. 
Improper handling of the binning and
interpolation issues typically results in errors at the few percent
level for all radii. The issue of extrapolating the integrals to infinity
usually only affects the recovered profiles at the largest measured
scales.

\subsection{Binning and Interpolation}
\label{section:binning}
Typically, one will construct a $\Delta\Sigma$ profile by
measuring the average weighted lensing distortion in a set of radial
bins. For example, \cite{sheldon:gmcf} measured $\Delta\Sigma(R)$ in
18 logarithmically-spaced radial bins from $0.02 ~h^{-1}$ Mpc to $11~
h^{-1}$ Mpc, giving a bin width $R_{i+1}/R_{i}=1.41$. One would like
to assign one effective radius to each bin. The radius assigned
should be that radius at which we believe the true $\Delta\Sigma(R)$
is equal to our measured average. The mean value theorem of integral
calculus guarantees that this radius will be inside the bin if
$\Delta\Sigma(R)$ is continuous, yet it is impossible to find this
correct value precisely without first knowing the shape of the function
within the bin.  In practice, however, it is good enough to approximate
the function as a local power-law, for which the correct radius can be
calculated exactly. The power-law exponent should roughly match the slope
of the data though the result is not very sensitive to the exact value.
This should be the first step carried out
before applying the density and mass formulas. For the binning and
data of Sheldon et al. this procedure produces 
only a $\sim 1$\% shift in effective radius from simply
using the midpoint, mean, or geometric mean of the bin boundaries $R_{i}$ and $R_{i+1}$.
For larger radial bins, say, as large
as $R_{i+1}/R_{i}=1.8$, this interpolation can lead to a 5\% shift in 
the assigned radius.

After selecting the proper effective radii, $R_{i}$, corresponding to
each $\Delta\Sigma(R_{i})$ bin, one can apply the inversion
formulae of \S 4. 
This requires a scheme for interpolating $\Delta\Sigma$
between the bins and for interpolating $\Delta\rho$ between the points
to infer the mass. \cite{saunders:2pt} applied Von Zeipel's inversion
formula, essentially our Equation \ref{eq:delta-rho}, to the projected
galaxy autocorrelation function, $w_p(R)$, to obtain the 3D
autocorrelation function, $\xi(r)$. They used a piecewise-linear model
for $w_p(R)$ and hence a piecewise-constant model for $w^{\prime}(R)$,
which enters into the Von Zeipel integral. This zero-th order
interpolation allows calculation of the integral over each bin
analytically, which results in simply a weighted sum over the data
points (details can be found in \cite{saunders:2pt}).
However, the data for both $w_p$ and $\Delta\Sigma$ look roughly like
power laws, $\Delta\Sigma \sim R^{-0.8}$, so this simple
interpolation scheme biases the inversion low by about 10\% for Sheldon et
al.-sized bins.  If one takes this scheme to first order by
representing the derivative, $\Delta\Sigma^{\prime}(R)$, as piece-wise
linear rather than constant over a bin, this biases the result by
about the same amount, but high instead of low.  One can in fact do
n-point Lagrangian interpolation to any order and still evaluate the
integrals analytically over each bin. Such methods are completely
linear in the data, which is another advantage.  In principle, higher
order interpolation will reduce this bias, but only at the expense of
correlating many neighboring points.  We have found in practice that
the best method is to use a power-law interpolation, equivalent to
linear interpolation of the log-log values. Any two positive data points 
define a unique power-law between them.
The integrals for each bin
need to be evaluated numerically, so this is slower computationally,
and care needs to be taken to handle negative data points which may be
present for noisy data. We therefore do not take logarithms directly
but rather model the function as a power law plus constant, where the
constant is only used for negative values. Data for which there are
negative values are generally very noisy; for that case,
interpolation is never the dominant source of errors.  Tests of the 
inversion show that for power-law density profiles, 
this interpolation scheme is exact
(unsurprisingly); for NFW profiles and halo-model type profiles, the 
power-law interpolation 
is good to a fraction of a percent for Sheldon et al.-sized bins
(i.e. $R_{i+1}/R_{i}=1.41$)
and very robust for noisy data.

\subsection{Endpoint Corrections}
\label{section:endpoint}
The inversion integrals in Sections 3 and 4 should in principle be
evaluated from the minimum radial bin, $R_{min}$, to $\infty$. In practice,
one only has shear data covering a range 
from $R_{min}$ to $R_{max}$. We refer to the
parts of the integrals from $R_{max}$ to $\infty$ as endpoint
corrections; they must be added to the parts of the integrals that
we can actually perform by interpolating the data. Consider 
Equation \ref{eq:delta-rho} as an example; we can rewrite it as 
\begin{equation}
\Delta\rho(r) = \frac{1}{\pi} \int_r^{R_{max}}dR~ \frac{-\Sigma^{\prime}(R)}{\sqrt{R^2-r^2}}
+ \frac{1}{\pi} \int_{R_{max}}^{\infty}dR~ \frac{-\Sigma^{\prime}(R)}{\sqrt{R^2-r^2}} \equiv T_D + T_C, 
\label{eq:drho-sep}
\end{equation}
where $T_D$ and $T_C$ stand for Data and Correction. 
To estimate the magnitude of each term, consider an isothermal
sphere, $\Delta\Sigma \sim R^{-1}$; in this case,  
the integrals can be computed analytically, leading to 
\begin{equation}
\frac{T_C}{T_D + T_C} = 1- \sqrt{1-\left( \frac{r}{R_{max}}\right)^2}.
\label{eq:TCTD}
\end{equation}
For the outermost data point, $r=R_{max}$, the correction term gives the
whole result, and one cannot learn anything about the density at
this point. However, for smaller scales, $r \ll R_{max}$, the
ratio $T_C/(T_D+T_C) \simeq (r/R_{max})^2/2$. For example, for 
Sheldon et al.-size
bins, this implies that the correction is only about 14\% at the value of 
$R_i=R_{16}$, two radial bins inward of $R_{max}=R_{18}$. 
The correction terms at the next bins inward (proceeding 
from larger to smaller scale) are 7\% , 3\%, and 2\%, while the 
remaining inner 12 bins are affected by less than 1\%.
In fact, it is only the error in estimating this correction term that
concerns us, so we expect that even at the second bin inward of 
$R_{max}$, one 
could estimate $\Delta \rho$ to an accuracy of a few percent. 
Around the virial radius of
clusters, $R \sim 1~ h^{-1}$ Mpc, the correction term itself 
is only half a percent of the total for 
$R_{max} =10~ h^{-1}$ Mpc. 

Beyond $10 h^{-1}$ Mpc, the halo density profile is expected to 
drop off more steeply
than isothermal, so the correction factor in this case should be less than
the estimate in \ref{eq:TCTD}. 
At these large scales, it is probably a good assumption to use
the shape of the linear mass correlation function (i.e. the Fourier transform
of the linear power spectrum), which is well understood
theoretically for a given cosmology, 
as the model with which to compute the endpoint
correction. This is the expected large scale behavior from the halo model.
This requirement of providing endpoint corrections is the
one part of this method which requires a model. 
However, this model
dependence only affects the results at the largest scales for any
reasonable extrapolation, guided by theoretical expectation and
constrained by the large-scale galaxy autocorrelation function which
is now measured to $175~ h^{-1}$ Mpc \citep{eisenstein:wiggles}. It is
left up to the experimenter to decide which points to omit and how to
take the error in the correction into account.

In this context, for lensing data that extends to scales of order a few Mpc, 
the endpoint correction can be considered a manifestation of the
mass sheet degeneracy. In the current paradigm of structure formation, 
the Universe is not filled with uniform-density mass sheets, but it 
does contain large-scale density perturbations that are correlated 
with dark matter halos. For a cluster positioned 
near the peak of a large-scale over-density with scale $\gg R_{max}$, 
the latter can be thought of to first approximation as a mass sheet 
for the cluster shear measurement. Using the theoretical mass 
correlation function on large scales to estimate the endpoint correction, 
as described above, therefore provides an estimate of the effective 
mass sheet degeneracy for this method. As we have argued above and 
show below in an N-body simulation, the effect of this 
degeneracy on the estimate of cluster virial masses is below 1\%.  

\subsection{Propagation of Errors}

In performing these inversions, it is important to correctly
propagate the errors from $\Delta\Sigma(R)$ to $\Delta\rho(r)$ and
$M(r)$. In particular, the covariance matrix between radial bins $R_i$
needs to be calculated. We will assume that we have already calculated
the covariance matrix for $\Delta\Sigma_{i}$ which we denote by
$C^{\Delta}_{ij}$. 
Usually this is just diagonal from shot noise (i.e., lensing
shape noise) but at larger scales there can be off-diagonal terms from
sample variance and systematics such as imperfect PSF anisotropy correction.
These off-diagonal terms can be estimated with jackknife resampling
\citep{sheldon:gmcf}.  We will denote the covariance matrix for
$\Delta\rho_i$ and $M_i$ as $C^{\rho}_{ij}$ and $C^{M}_{ij}$.

As we have mentioned, with the simplest polynomial interpolations, 
the mapping
$\Delta\Sigma(R) \longmapsto \Delta\rho(r)$ is linear, so there exists a
matrix $\bf{A}$ such that $\Delta\rho_{i}= A_{ij} \Delta\Sigma_{j}$; likewise,
since $M$ is a linear combination of $\Delta\Sigma$ and $\Delta\rho$, and
$\Delta\rho$ is computed linearly from $\Delta\Sigma$,
there exists a matrix $\bf{B}$ such that $M_{i}= B_{ij} \Delta\Sigma_{j}$.
For the polynomial interpolation schemes, these
matrix elements can be computed analytically, 
without requiring numerical integration.
For the power-law interpolation that we recommend using, the mapping is 
not \emph{exactly} linear, but 
it is well approximated by its first order Taylor expansion, 
\begin{equation}
A_{ij} = \frac{\partial \Delta\rho_i}{\partial \Delta\Sigma_j} ~~;~~~
B_{ij} = \frac{\partial M_i}{\partial \Delta\Sigma_j}.
\end{equation}
Although, these derivatives can be calculated analytically in principle, 
in practice we estimate them
numerically with a finite difference method. 

Alternatively, one could use the linear interpolations
just for the covariance matrices, where the matrix elements 
can be computed analytically. This would be the
fastest method computationally. Once the linearization has been computed (by any method), 
the covariance matrices can be propagated simply with matrix multiplication, 
$\bf{C^{\rho}}=\bf{A}~\bf{C^{\Delta}}~\bf{A}^T$ 
and $\bf{C^{M}}=\bf{B}~\bf{C^{\Delta}}~\bf{B}^T$.
Generally, we find that the covariance matrix $\bf{C^{\rho}}$ is mostly diagonal if $\bf{C^{\Delta}}$ is diagonal.
The covariance matrix $\bf{C^{M}}$, however, has large off-diagonal terms since $M(r)$ is a cumulative
statistic and therefore neighboring bins are correlated.

\section{Tests of the Method: N-body Simulations}
\label{section:tests}

As a proof of principle, we have performed tests of these
inversion methods on an N-body CDM simulation. In the simulation, we
can measure the 3D density and mass profiles directly and check that 
the inversion of projected quantities correctly recovers the
true values.

The simulation we use has $512^3$ particles
in a periodic cube of length 300 $h^{-1}$ Mpc. The simulation is
evolved from $z=60$ to $z=0$ using a TreePM code
\citep{white:mass-function,white:planck} 
\footnote{The simulation is publicly available at http://mwhite.berkeley.edu/Sim1/}
; we use only the $z=0$
output. The cosmological parameters used are $\Omega_M=0.3$,
$\Omega_{\Lambda}=0.7$, $h=0.7$, $n=1$, $\Omega_bh^2 =0.02$, and
$\sigma_8=1$.  The simulation has an effective Plummer force-softening
scale of 20 $h^{-1}$ kpc which is fixed in comoving coordinates.  The
mass of each dark matter particle is $1.7 \times 10^{10} h^{-1}
M_{\sun}$.  Dark matter halos are identified using a
Friends-of-Friends (FoF) algorithm \citep{davis:fof} with a linking
length of 0.2 in units of the mean inter-particle separation. Specific
details of the simulation such as resolution, cosmology, and halo
finding are not crucially important, since we are only interested in
whether the inversion methods recover the 3D quantities. The
simulation box was chosen to have high enough resolution to resolve
the inner regions of clusters measurable by SDSS and is large enough to 
have relatively low 
cosmic variance; nevertheless, it is quite a bit smaller than the size 
of the final SDSS cluster sample currently in preparation.

We select all halos of mass $M_{vir} > 10^{14} h^{-1}$ Mpc,  
measure $\rho(r)$ and $M(r)$, and average these 3D 
quantities for all such massive halos. For $\rho(r)$, we correct for
the effects of binning in a similar way to the methods described in
Section \ref{section:binning}. Next, we project the box separately along each
of its three axes (x,y,z). For each of these projections,
we measure the average $\Sigma(R)$ and $\Delta\Sigma(R)$ from Equation
\ref{eq:delta-sigma}.  We do not perform ray tracing to determine the
real shear, so we are only testing the inversion method, assuming that
one can measure $\Delta\Sigma$ through measurement of background galaxy 
shear (see Sheldon, et al.). In particular, we are 
not testing the shear
non-linearities, the details of galaxy shape measurement, photometric redhshift
estimation and calibration, or any kind of cluster finding. 
These details are important for any analysis of real data, but they 
are beyond the scope of this paper (and will be discussed further in 
future papers in this series).

By studying the differences in these projected 2D quantities for the
three different projections of the simulation box, 
we can crudely measure a kind of sample variance
which we call anisotropic sample variance (ASV). If the halo-mass
correlation function were exactly isotropic, all three projections
would agree precisely. Since this is only true in the infinite volume
limit, there will be differences between the three projections of the 
simulation as well as small departures from isotropy 
in any real measurement.  On small scales, these
differences arise from the asphericity of halos: if halo
orientations are not strongly correlated, the residual ellipticity
will result in a small random error that will decrease as 
$\sim 1/\sqrt{N_{halos}}$. More explicitly, we can define the halo 
ellipticity
along the $z$-direction as $e_z = (3 \sigma_z^2 - \sigma^2)/\sigma^2$, 
where $\sigma_z^2$ is the second moment of the
cluster density distribution  
along the projected direction and $\sigma^2 = \sigma_x^2 +
\sigma_y^2 +\sigma_z^2$ defines the characteristic cluster size.

The ensemble average of $e_z$ is zero by symmetry, but the rms for a finite
sample will be zero within $\pm \sqrt{<e_z^2>/N_{halos}}$.  One can
show that this rms halo ellipticity induces a multiplicative correction 
factor in the projected density and all
derived quantities ($\rho$ and M) of $1 \pm
\sqrt{<e_z^2>/N_{halos}}$; this factor will vary randomly from realization 
to realization (or from projection to projection within a single 
N-body realization). The rms halo ellipticity $\sqrt{<e_z^2>}$
appears to be about 0.5 for the massive halos in the N-body 
simulation described above.  Since large survey samples will 
have very many clusters, this correction will usually be a very
small effect.  For example, in the sample used in this study, there
are 1226 halos above the mass threshold in the simulation, 
resulting in an expected multiplicative variance of $\pm
1.4\%$ in the inferred profiles. 
This estimate agrees very well with the observed scatter between the
three different projections on small and intermediate scales, as shown below in Figures 
2 and 3. The SDSS cluster samples are typically larger than this, so 
the effect will be correspondingly reduced. Less massive halos (these are
$M=10^{14} h^{-1}M_{\sun}$), being more common, will have a larger 
$N_{halos}$ and thus this will be reduced further. 

On larger scales, where the halo-mass correlation function is dominated 
by contributions from mass elements in halos different from the lens 
halo (i.e., where, in halo model terms, the two-halo term 
dominates \citep{seljak:halos,mandelbaum:diss-v-halo}), 
the asymmetric sample variance 
arises predominantly from the shapes of larger-scale structures such as
filaments and superclusters. Since there are fewer structures contributing to this part of the 
ASV (fewer filaments than halos), and since these large structures are more 
asymmetric than the halos, the variance between projections 
is expected to be larger on large scales than small scales, in 
qualitative agreement with 
Figure \ref{fig:rho} below.
This large-scale asymmetric sample
variance is expected to scale as $1/\sqrt{V}$ where $V$ is the volume
of the survey. When the shot noise from galaxy shapes (i.e., the shape
noise) is sufficiently small, the large-scale errors will be dominated
by the ASV. Judging by the volume of our simulation and the scatter
that we see at scales of a few Mpc/h, we expect this large-scale multiplicative
error to scale roughly as $2\% ~\times (V/Gpc^3)^{-0.5}$, although it
may vary with scale $R$.

\begin{figure}
\epsscale{0.9}
\plotone{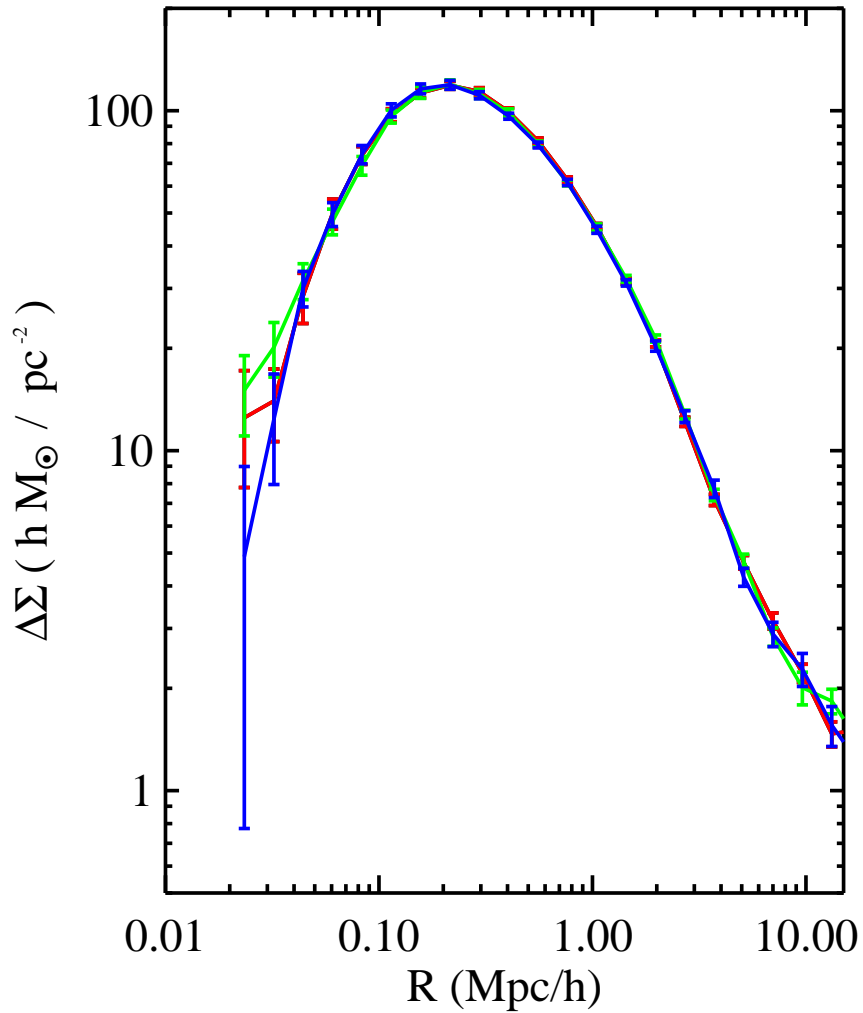}
\caption{
The mean tangential shear $\Delta\Sigma(R)$ for halos with virial masses 
greater than $10^{14} h^{-1} M_{\sun}$ for three orthogonal 
projections of a $\Lambda$CDM N-body simulation. The turn-over below 
$R \sim 20~h^{-1}$ kpc is due to the finite resolution of the simulation. 
The scatter between the curves at the smallest scale is likely due to shot noise since
there are very few dark matter particles at these scales due to the resolution. 
At large radius, we attribute 
the differences to anisotropic
sample variance (ASV) caused by large-scale structure along the line of sight.
\label{fig:ds}}
\end{figure}

Figure \ref{fig:ds} shows the $\Delta\Sigma$ results for each of the
three projections of the simulation (the three different colors 
correspond to the three different projections). We note that these curves turn
over below about $20 h^{-1}$ kpc, an effect almost certainly 
due to the finite resolution of the simulation: this scale is 
about 10 times the Plummer
force-softening scale.  The turn-over in $\Delta\Sigma$
typically occurs at a few times the 3D correlation function core
radius and the 3D correlation function core radius typically occurs at
a few times the Plummer force softening scale. By correlation function core
radius we have in mind something like a power-law with a softening,
$r \rightarrow \sqrt{r^2+r_{core}^2}$.
Smaller-volume,
higher-resolution simulations do not show this turn-over in
$\Delta\Sigma$. For a pure NFW profile, $\Delta\Sigma$ flattens at
small scale but does not decline.  As noted above, despite
this resolution issue, we should be able to use these small scales to
test our inversion methods since the 3D values are equally affected by
resolution. In fact, this gives us an opportunity to test the 
robustness of the methods by including a pathological regime instead of 
a perfect power-law or NFW profile on all scales.  
The large errors (estimated by a jackknife technique) at
small scales indicate that there are not many particles in these
weak cusps. The larger deviations at the largest scales indicate the
larger ASV, as expected from the arguments above.

We invert each of these $\Delta\Sigma$ profiles to obtain the estimated 
density $\Delta\rho(r)$ and $M(r)$ using the formulae and algorithms
described in previous sections. The endpoint corrections are carried out 
assuming power-law extrapolations with logarithmic slopes 
that are not too different from the
expected continuation of the linear theory correlation function.  The last
two bins from the inversions are not plotted, since they are highly
affected by the endpoint corrections.

Figure \ref{fig:rho} shows results of the inversion for the 
density profile $\rho(r)$. 
As in Fig. 1, the red, green, and blue
curves denote the inversions from the three different projections.  The
magenta line shows the true 3D density as measured in spherical
shells. On this log-log plot one can hardly see any difference between
the true and inverted profiles. 
The bottom panel of Figure \ref{fig:rho} 
shows the ratio of each of these inverted density profiles 
to the true 3D density with a linear axis. On scales
$r <1~ h^{-1}$ Mpc, these ratios are all consistent with unity to
within about 5\%. On larger scales, the ASV becomes larger, and the
three curves differ randomly from unity by about 20\%. There
do not appear to be any systematic biases or significant
trends. Changing the endpoint correction moves these curves around
by a few percent at the largest scales.

Figure \ref{fig:mass} shows results of the inversions for the mass
profile, $M(r)$. Again, the top panel shows all three inversions with
the true $M(r)$ over-plotted in magenta; the inversions and the 
true profile are indistinguishable
on this plot. The lower panel shows the ratios of inverted to true 
mass. Except for the
smallest scales, where the scatter between the three inversions is a
little larger, the mass ratios are consistent with unity to about
5\%. Around the virial radius, $r \sim 1 h^{-1}$ Mpc, which is the optimal
location 
in terms of measurement signal-to-noise, the inverted 
masses are correct to about
2\%. There appears to exist a slight tilt to these ratio curves, but the
points at different radii are strongly 
correlated ($M(r)$ being a cumulative quantity), so this trend
is not significant.  On the whole, we consider these tests a
confirmation that inverting lensing measurements can correctly recover
the density and mass profiles of galaxy and cluster halos.

\begin{figure}
\epsscale{0.75}
\plotone{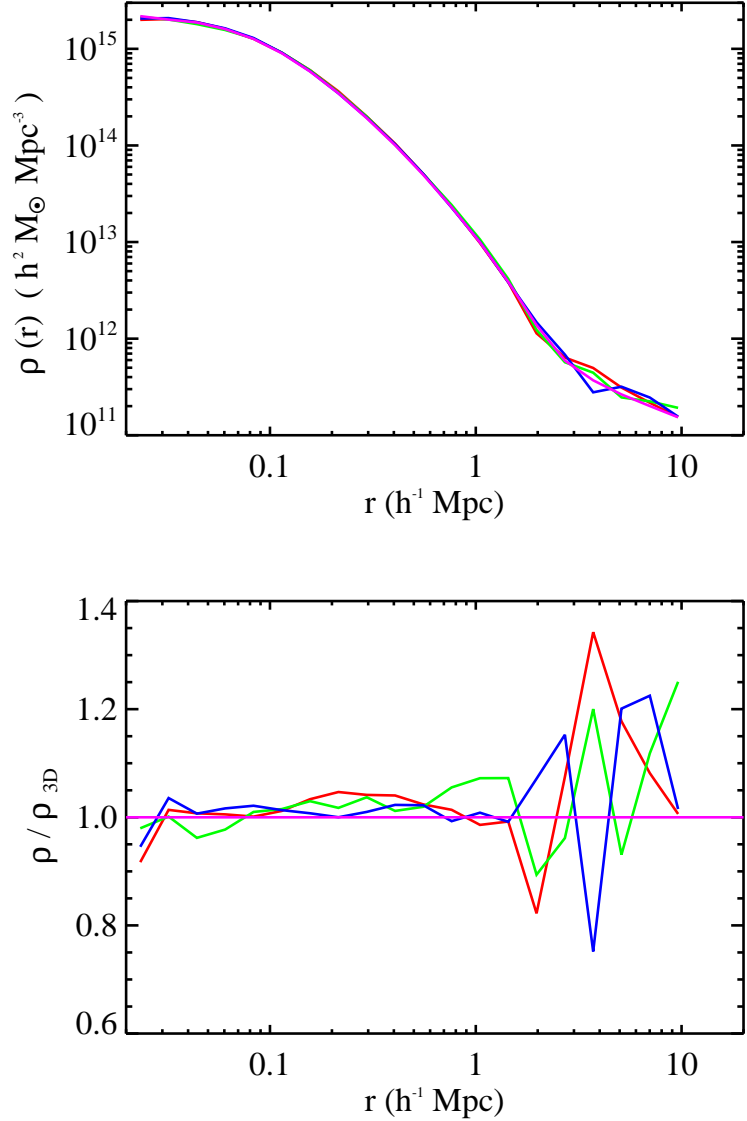}
\caption{
Top panel: Inverted 3D density profiles for the three projections of the 
simulation.
The red , green, and blue curves indicate the inversions, while the 
magenta curve denotes the true density profile measured 
by averaging the dark matter particle density in spherical shells. 
The inverted profiles are only distinguishable from each other and from 
the true profile at large scales.
Bottom panel: ratio of the inverted density to the true 3D density for 
the three simulation projections as a function of radius; curves correspond 
to the same projections as in the top panel. For each projection, 
the inverted and true density profiles agree to about 5\% or better 
within the virial radius ($r_{200}=0.585 ~h^{-1}$ Mpc). On larger scales, there is significantly
more scatter due to ASV in this relatively small simulation box.
\label{fig:rho}}
\end{figure}

\begin{figure}
\epsscale{0.75}
\plotone{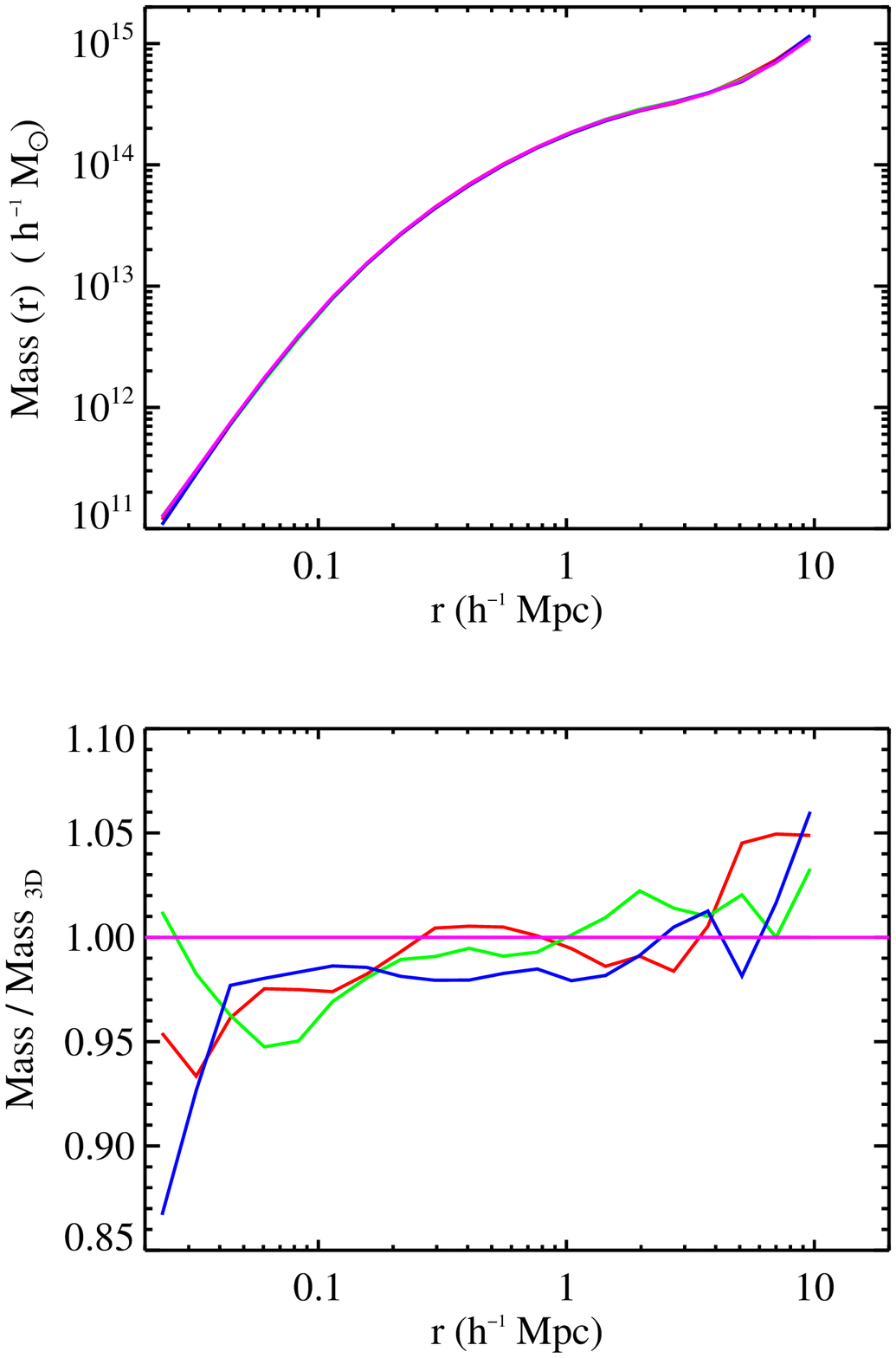}
\caption{ Top panel: mass
profiles inverted from the tangential shear 
for the three projections of the N-body simulation 
(red, green, and blue curves). The magenta line shows the true average mass
profile measured in 3D for these halos. The differences are not evident on 
this scale. Bottom panel: ratios of the three inverted 
mass profiles to the
true 3D profile. The inverted masses are all correct to within 5\% out 
to the largest scales and to within 
2\% near the virial radius. Because the points are correlated, the
slight trend with radius that is apparent is not significant. The deviations
at the smallest scale are mostly likely due to shot noise.
\label{fig:mass}}
\end{figure}

\section{Effects of Halo 
Asphericity and Large-scale Structure along the Line of Sight}
\label{section:lss-los}

In the case of weak lensing mass reconstructions for individual clusters,
one must consider the systematic 
effects of both cluster asphericity and contamination from structure along
the line of sight, such as filaments or clusters at other distances seen 
in projection. Several studies
\citep{cen:projection-effects,metzler:cluster-lss-contam,metzler:lss-contam,white:clusters-completeness} 
have concluded that large-scale structure (LSS) along the line of sight
introduces significant scatter and bias in the resulting cluster 
mass estimates. Later studies \citep{hoekstra:lss-contam1,hoekstra:lss-contam,
clowe:lensing-proj,deputter-white} 
agreed that LSS introduces scatter in the mass estimate, 
though \cite{clowe:lensing-proj} argued that the bias found in the
previous studies was due to an over-simplification of the mass
estimation.  \cite{dodelson:lss-contam} suggested ways that the extra
noise from LSS can be reduced, though not eliminated, through a
particular type of filtering when 
making mass maps.  \cite{clowe:lensing-proj}, on the other 
hand, argued that
cluster halo asphericity  is a larger problem for
determining cluster masses than LSS along the line of sight: if the
major (minor) axis of the halo 
is aligned along the line of sight, the virial mass will be
overestimated (underestimated). In general one does not
know the orientation of the halo in 3D, so this creates an 
uncertainty in the inferred mass at the $\sim$
30\% level. 

As demonstrated in Section \ref{section:tests} and Figure \ref{fig:mass}, 
neither of these effects 
compromises the accuracy of our statistical mass inversion method, 
provided the cluster sample is sufficiently large.  
By stacking many clusters to form a statistical sample, the effects of 
large-scale structure uncorrelated with the lensing cluster must average out 
by statistical isotropy. On small scales, the mean orientation 
of aspherical halos for a large sample of lenses must be very close to random, provided 
that the method of {\it selecting} cluster lenses is not 
biased toward including those oriented, say, along the line of sight. 
These selection effects can be checked in N-body simulations that 
incorporate some prescription for assigning luminous galaxies to dark 
matter halos; as such, they are beyond the scope of this paper, but 
we plan to address them in a future publication. (Note that cluster 
samples selected by their shear signal may suffer from such a bias 
\citep{white:clusters-completeness}.) On intermediate scales, 
the orientation of nearby filamentary structure correlated with the cluster 
lens (as well as the angular distribution of nearby correlated halos) must 
also be random for a large statistical sample. Since such structure is 
correlated with the lens, it {\it does} contribute to the tangential shear 
on large scales; indeed, as we have argued above, it dominates the 
signal from clusters on scales above a few $h^{-1}$Mpc and simply represents the 
large-scale cluster-mass correlation function. Again, as shown above, 
the effects of correlated structure along the line of sight appear to be 
negligible out to the virial radius of massive clusters.

The only residual problem for the statistical inversion method is 
anisotropic sample variance at large scales 
(discussed above in Section \ref{section:tests}); this finite-volume effect  
creates a small scatter in the inferred mass 
but does not create a bias. 

\section{Discussion}

We have developed a non-parametric method for inverting cross-correlation 
lensing measurements to
obtain average density and mass profiles of halos. 
We have demonstrated the method on an N-body simulation and shown that
it successfully recovers the 3D profiles. 
We argued that asphericity of halos and large-scale structure
along the line of sight do not introduce bias or substantial 
uncertainty in the inferred mass estimates.

This method should find several useful 
applications in surveys. 
Applied to galaxy-galaxy lensing measurements, this method can be
used to measure $\Delta\rho(r) = \Omega_m\rho_{crit}~\xi_{gm}(r)$ for 
samples of lens galaxies (e.g., Sheldon, et al. 2004).
On small scales, this provides information about galaxy dark matter halos
and should thereby constrain models of 
galaxy formation and evolution, including 
the details of hierarchical structure formation and merging. 
Statistical mass profiles around galaxies can provide
virial mass measurements and, combined with the average light profile
around these galaxies, determine mass-to-light ratios as a function of scale.

On larger scales, galaxy-galaxy lensing inversions should tell
us more about cosmology.  The autocorrelation function of galaxies
allows us to measure $\xi_{gg}(r) = b^2(r) \xi_{mm}(r)$, where $b(r)$
is the scale-dependent bias and $\xi_{mm}(r)$ is the mass
autocorrelation function.  Lensing allows us to measure
$\Omega_m~\xi_{gm}(r) =
\Omega_m~b(r)~\mbox{r}_{\times}(r)~\xi_{mm}(r)$, where the cross-bias,
$\mbox{r}_{\times}(r)$, is sometimes referred to as stochastic bias (a
term we do not recommend using in this context, since it is not
bounded by $\pm1$). The presence of these two bias functions, $b(r)$
and $\mbox{r}_{\times}(r)$, allows for the most general model of the
relative clustering of galaxies and mass. 
(See also \cite{neyrinck:halo-model}.)
Simulations indicate that $\mbox{r}_{\times}(r)$ is consistent with 
unity on scales larger than $\sim 2 h^{-1}$ Mpc for any range of halo masses
\citep{taz:dissipationless}. 
In this case, measurement of 
large-scale galaxy-galaxy lensing and of the 
galaxy autocorrelation function 
determine the mass correlation function, $\Omega_m^2 \xi_{mm} = (\Omega_m
\xi_{gm})^2/\xi_{gg}$ and so fixing the \emph{shape} of the linear correlation function
determines the amplitude or normaliztion $\Omega_m\sigma_8$. 
It also determines the bias, $b(r)/\Omega_m =
\xi_{gg}/\Omega_m\xi_{gm}$.  The assumption that bias is scale
independent on large scales is crucial to extracting cosmological
information from the galaxy power spectrum; lensing
can provide a way to test this important assumption. 
Other cosmological probes, such as cluster 
counts and cosmic shear, measure a different parameter combination,
$\Omega_m^{0.5} \sigma_8$, so combining these probes with
cross-correlation lensing helps break this classic
degeneracy. An advantage of this new method is that it is robust;  
the only assumptions used are that general relativity is correct
and that $\mbox{r}_{\times}(r)\simeq 1$ on large scales.

Applied to galaxy clusters, this inversion method should prove useful for 
both astrophysics and cosmology.  On small scales, cross-correlation
lensing probes the mean density profiles of clusters. Cosmological dark 
matter simulations
indicate that clusters have universal dark matter halos (though
baryonic physics presumably 
needs to be taken into account to understand the inner density 
structure in detail). 
Inverted density profiles can test this assumption quite directly. As we have
demonstrated, cluster mass profiles can be determined, leading to 
virial mass estimates independent of a model for
the density profiles. Virial masses allow a direct comparison between
simulated and real clusters. In addition, in a cluster survey 
we can measure the cluster 
abundance, $n(M,z)$, 
binned by any observable proxy for mass, and use this inversion method to 
calibrate the virial mass-observable relation as a function of redshift. 
This should allow cluster surveys to more precisely probe cosmology, including 
the dark energy. We still need to gather information about the scatter in 
the mass-observable relation; this can be constrained by simulations, by 
self-calibration \citep{limahu2}, and internally by dividing the clusters into 
subsamples and separately estimating the mass-observable relation. 

Before applying this method with confidence to real cluster data, the possible 
selection biases involved in a given cluster {\it selection} algorithm 
need to be explored. If the selection algorithm preferentially finds  
clusters aligned along the
line of sight, the assumption of statistical isotropy 
will be violated and the mass inversions
may be biased. Similarly, if the selection algorithm assigns cluster centroids 
that are displaced from the true halo centers, the resulting 
density profiles will be convolved with distribution of centroid errors. 
Additional complications for the cluster counting technique 
arise if clusters, however selected, are not 
isomorphic to massive dark matter halos. 
These issues can be tested with simulations. 

Applying these lens inversion methods to clusters on large scales provides 
new cosmological information beyond the cluster counting technique.  
As noted for galaxies above, we can use the cluster-mass and cluster-cluster 
correlations to 
measure $\Omega_m\sigma_8$. A useful cross-check is provided by comparing 
the results for different cluster samples binned by some observable, 
e.g., richness. This method should be even more robust for clusters 
than for galaxies, since the halo bias as a function of mass is predictable
from theory and can be studies in dissipationless N-body simulations 
\citep{seljak-warren:halo-bias,taz:dissipationless}.  One can use this information in two ways. 
Conservatively, one could measure the average cluster
masses, $M$, and $b(M)/\Omega_m$; from these bias measurements one could
further constrain cosmology through the theoretical bias predictions
and the direct measurement of $\Omega_m\sigma_8$.  Alternatively, one 
could use the lensing data only on small scales to determine masses
and assume that the theoretical bias predictions are correct; in this 
approach, one would use the mass and an assumed cosmology to predict
the bias. This would allow one to `de-bias' the cluster autocorrelation
function and obtain a constraint on $\sigma_8$ that depends on
other cosmological parameters as well.  Since the scaling of bias
with mass is somewhat weak ($b(M) \sim M^{0.2}$ in the vicinity of
$M=10^{13} h^{-1}M_{\sun}$), and the cluster autocorrelation function
can be measured with high signal-to-noise, this measurement of
$\Omega_m\sigma_8$ could be quite precise.  This method has been
applied to galaxies by \cite{seljak:bias}, utilizing the halo model. 
We believe that clusters are a more natural choice for this method in that
they do not need an additional halo occupancy presciption.

The approach of `de-biasing' the correlation function also makes
possible the study of the growth of structure. In principle, the inversion 
method can be applied to data in redshift bins, 
allowing one to measure the large-scale 
mass power spectrum amplitude as a function of 
redshift. This is particularly useful, because
the linear density perturbation growth rate is sensitive
to dark energy. Moreover, since the evolution of the halo 
mass function is
also sensitive to the growth of structure, combining cluster counts 
and the cluster-mass correlation function should offer a powerful 
probe of cosmology.

We are currently applying these statistical 
lensing inversion techniques to optically selected 
clusters in the Sloan Digital Sky Survey. In the future, on-going 
and planned wide-area surveys, including RCS II, CFHTLS, the VST 
and VISTA surveys,  
the Dark Energy Survey, PanSTARRS, LSST, and SNAP/JDEM 
will allow for higher signal-to-noise lensing measurements to be
made and will eventually probe much larger cosmological volumes. 
The cosmological constraints they supply from cross-correlation 
lensing will complement those from other probes.

\section{Acknowledgments}
DEJ would like to thank the KICP for its
hospitality on many visits as well as the Aspen Center for Physics.  
We would like to thank Martin White and
Andreas Berlind for providing the N-body simulation that was used in this
paper. RHW was supported by NASA through Hubble Fellowship
HF-01168.01-A awarded by Space Telescope Science Institute.

\begin{singlespace}
%\bibliography{biblio}
%\bibliographystyle{apj}
%\input{bibliography.tex}

\end{singlespace}
\end{document}